\documentclass[sigconf]{acmart}
\AtBeginDocument{%
  }

\usepackage{subfigure}
\usepackage[compact]{titlesec}
\usepackage{setspace}
\titlespacing{\section}{0em}{0em}{0em}
\titlespacing{\subsection}{0em}{0em}{0em}
\titlespacing{\subsubsection}{0em}{0em}{0em}
\titlespacing{\paragraph}{0em}{0em}{0em}
\setcopyright{acmlicensed}
\copyrightyear{2018}
\acmYear{2018}
\acmDOI{XXXXXXX.XXXXXXX}
\acmConference[Conference acronym 'XX]{Make sure to enter the correct
  conference title from your rights confirmation email}{June 03--05,
  2018}{Woodstock, NY}
\acmISBN{978-1-4503-XXXX-X/2018/06}

\usepackage{subcaption}
\usepackage{enumitem}
\setlist[itemize]{noitemsep, topsep=0pt, partopsep=0pt,left=0pt}
\setlist[enumerate]{noitemsep, topsep=0pt, partopsep=0pt,left=0pt}

\begin{document}

\title{Scalable GPU Performance Variability Analysis framework}

\author{Ankur Lahiry}
\affiliation{%
\institution{Texas State University}
\city{San Marcos, Texas}
\country{USA}}
\email{vty8@txstate.edu}

\author{Ayush Pokharel}
\affiliation{%
\institution{Texas State University}
\city{San Marcos, Texas}
\country{USA}}
\email{ssu22@txstate.edu}

\author{Seth Ockerman}
\affiliation{%
\institution{University of Wisconsin-Madison}
\city{Madison, Wisconsin}
\country{USA}}
\email{sockerman@cs.wisc.edu}

\author{Amal Gueroudji}
\affiliation{%
\institution{Argonne National Laboratory}
\city{Lemont, Illinois}
\country{USA}}
\email{agueroudji@anl.gov}

\author{Line Pouchard}
\affiliation{%
\institution{Sandia National Laboratories}
\city{Livermore, California}
\country{USA}}
\email{lcpouch@sandia.gov}

\author{Tanzima Z. Islam}
\affiliation{%
\institution{Texas State University}
\city{San Marcos, Texas}
\country{USA}}
\email{tanzima@txstate.edu}

\renewcommand{\shortauthors}{Lahiry et al.}

\newcommand{\ankur}[1]{\textcolor{blue}{#1}\xspace}
\newcommand{\ayush}[1]{\textcolor{orange}{#1}\xspace}
\newcommand{\todo}[1]{\textcolor{red}{TODO: #1}\xspace}

\begin{abstract}
  Analyzing large-scale performance logs from GPU profilers often requires terabytes of memory and hours of runtime, even for basic summaries. These constraints prevent timely insight and hinder the integration of performance analytics into automated workflows. Existing analysis tools typically process data sequentially, making them ill-suited for HPC workflows with growing trace complexity and volume. We introduce a distributed data analysis framework that scales with dataset size and compute availability. Rather than treating the dataset as a single entity, our system partitions it into independently analyzable shards and processes them concurrently across MPI ranks. This design reduces per-node memory pressure, avoids central bottlenecks, and enables low-latency exploration of high-dimensional trace data. We apply the framework to end-to-end Nsight Compute traces from real HPC and AI workloads, demonstrate its ability to diagnose performance variability, and uncover the impact of memory transfer latency on GPU kernel behavior.
\end{abstract}



\keywords{Do, Not, Us, This, Code, Put, the, Correct, Terms, for,
  Your, Paper}


\maketitle

\section{Introduction}


GPUs are critical accelerators for High-Performance Computing (HPC) workloads due to their parallel processing capabilities. Efficient profiling of GPU-driven applications is essential to identify performance bottlenecks, manage resources, and optimize GPU utilization. However, advanced profiling tools generate performance logs that rapidly grow to terabytes, overwhelming traditional sequential analysis methods that load entire datasets into single-node memory. Current approaches rely heavily on incremental processing or repeated disk-based operations, severely limiting scalability and efficiency.

To overcome these limitations, we propose a distributed, data-driven analysis framework designed specifically for GPU performance logs. Our solution partitions large SQLite3 tables into smaller shards processed concurrently across multiple MPI ranks. This significantly reduces per-node memory demands and computational latency by allowing each node to analyze only a fraction of the data, thereby enabling scalable, efficient analysis of massive GPU datasets.
\begin{figure*}[t]
    \centering
    \subfigure[Variation in memory stall across all ranks]{
    \includegraphics[width=0.38\textwidth]{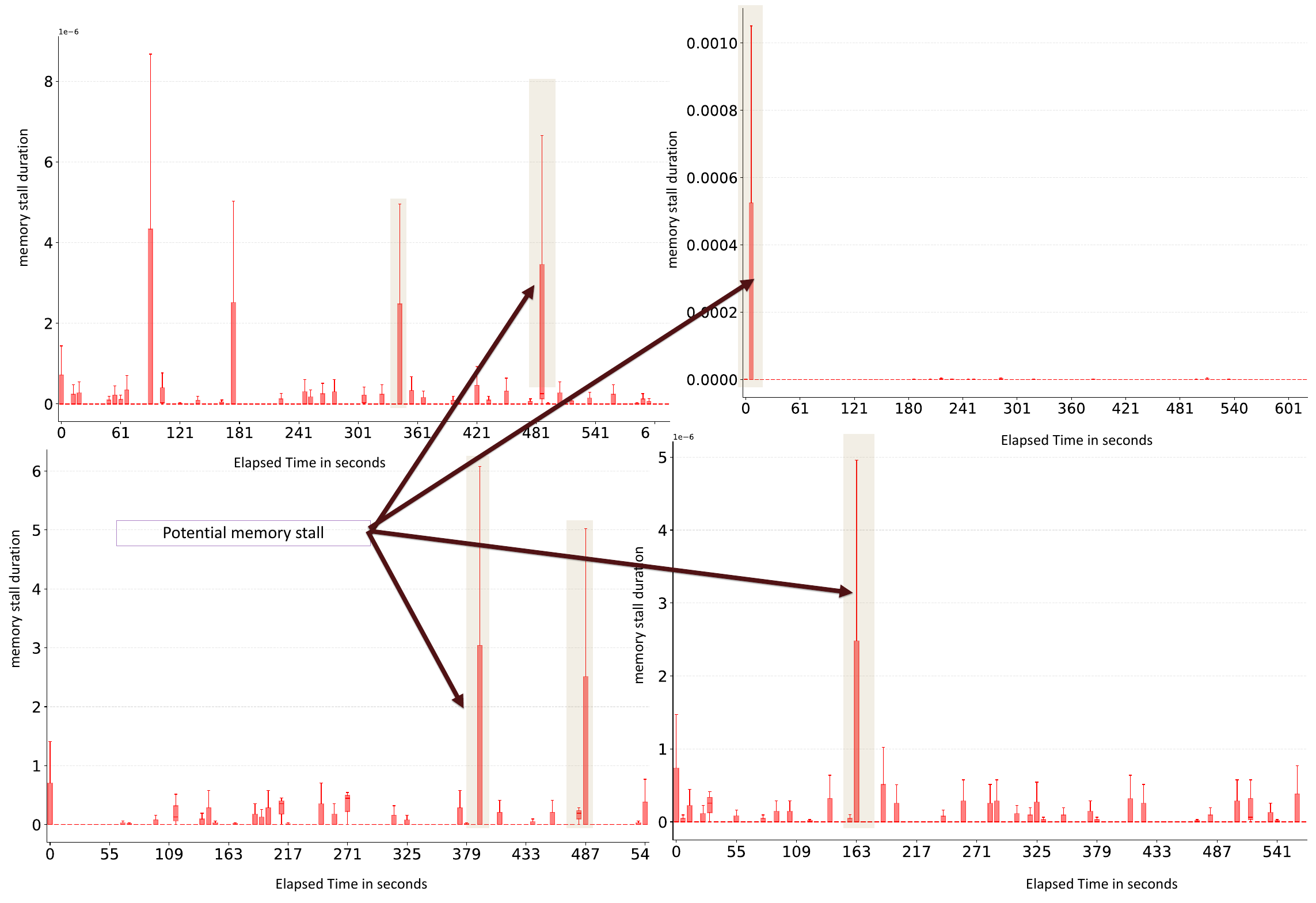}
        \label{fig:memory_stall_ranks}
        }
    \subfigure[Root causes analysis for Rank 2]{
    \includegraphics[width=0.38\textwidth]{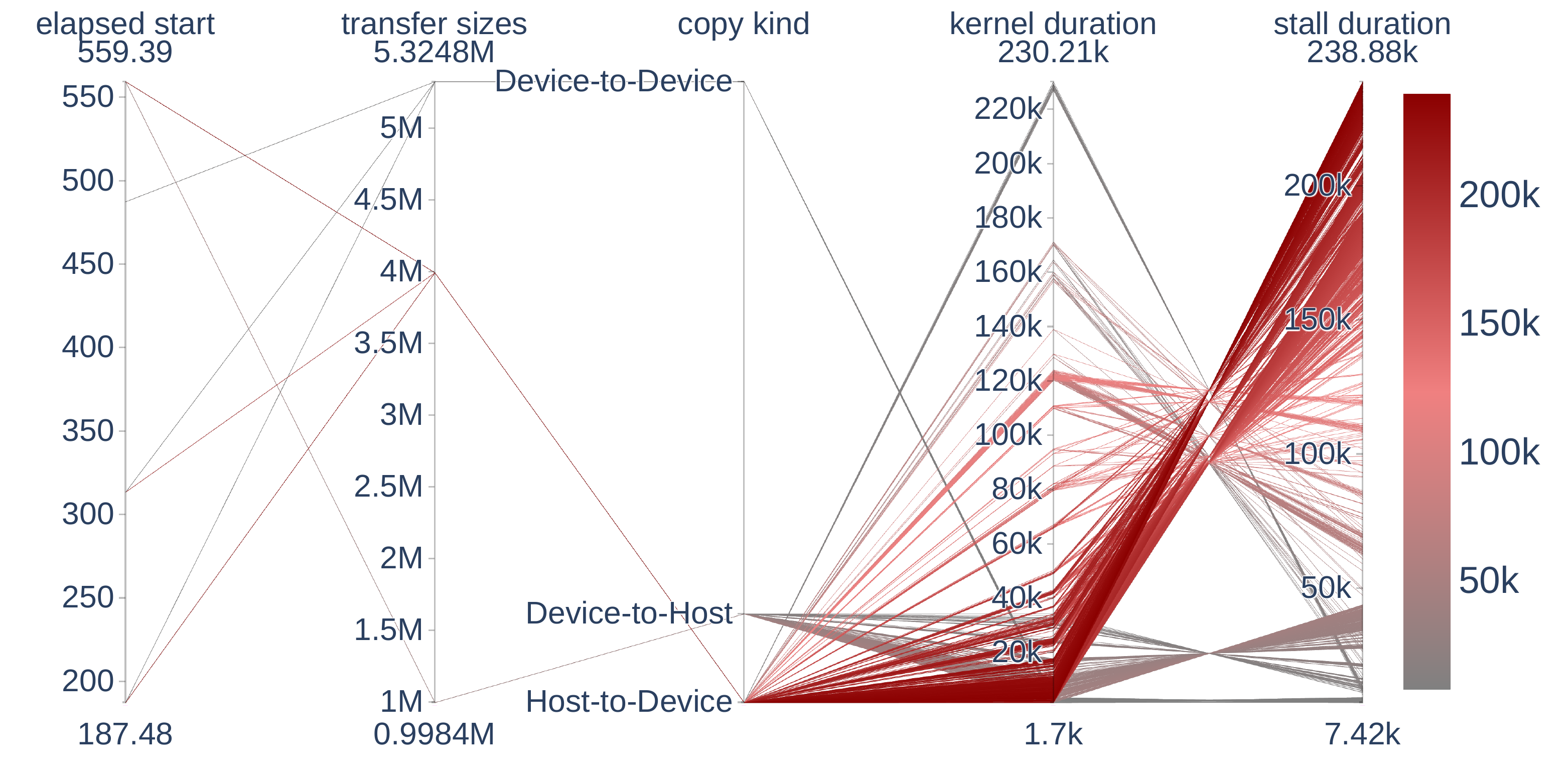}
            \label{fig:memory_stall}
            }
    \subfigure[Overhead]{
    \includegraphics[width=0.2\textwidth]{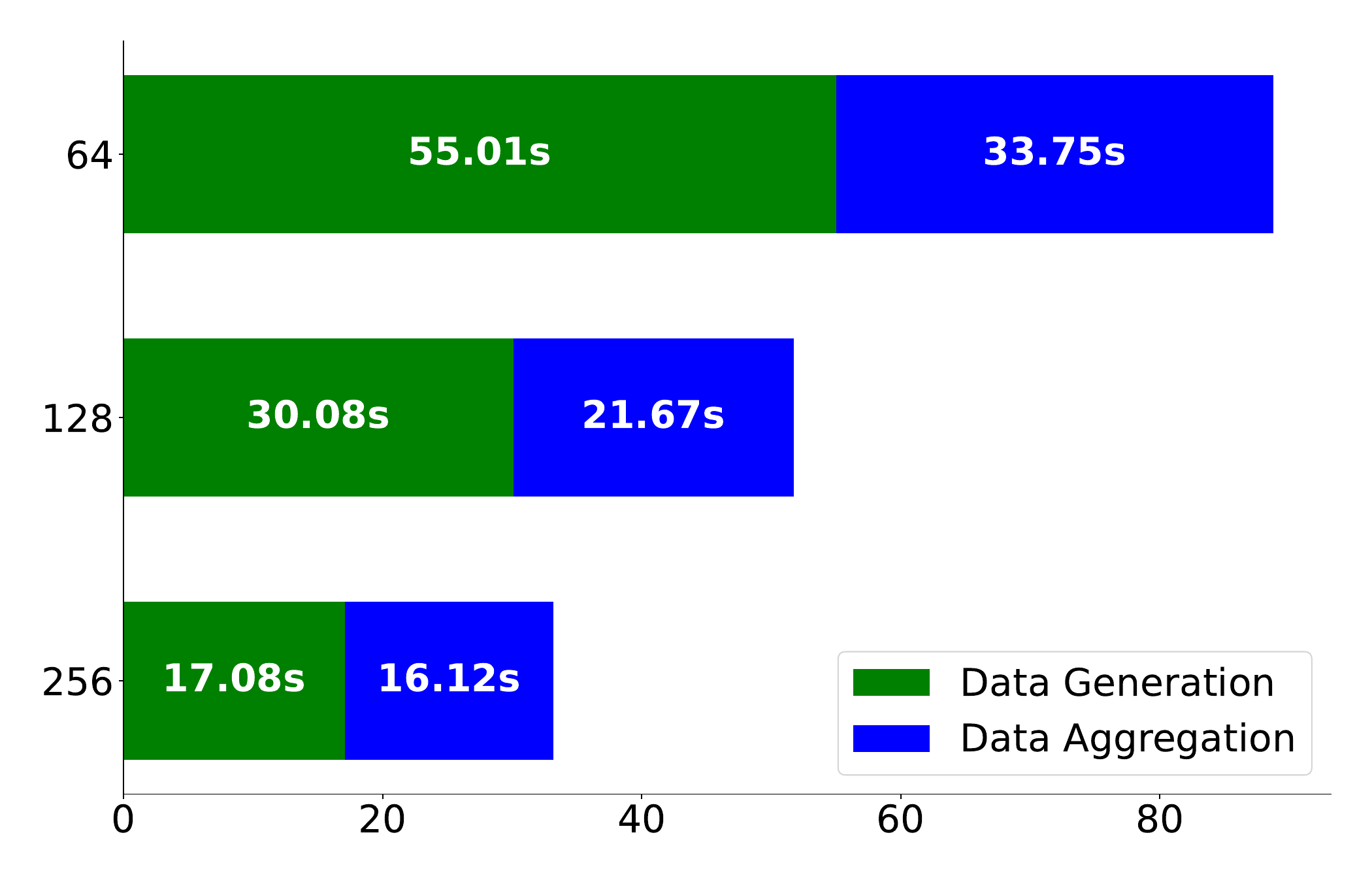}
    \label{fig:timing}
    }
    \caption{Comparison of memory stall metrics: (a) all four ranks, (b) Rank 2 parallel coordinate plot, and (c) overhead of the pipeline.}
    \label{fig:memory_stall_combined}
    \label{-0.2in}
\end{figure*}
\section{Background}

\subsection{Nvidia Nsight Profiling}
\label{sec:profiling_bg}
NVIDIA Nsight Profiling (Nsight Systems and Nsight Compute) traces CPU-GPU interactions, kernel executions, memory transfers, and hardware usage. Profiling data is organized into CUPTI tables (Table~\ref{tab:table_for_profiling}), using the dataset from Gueroudji et al.~\cite{gueroudji2024performance}. In the dataset, (1) the 
ACTIVITY\_KIND\_KERNEL records kernel launches, timestamps, device and stream IDs, and resource usage; (2) 
ACTIVITY\_KIND\_MEMCPY logs memory transfers, timestamps, sizes, directions, and stream IDs; (3) TARGET\_INFO\_GPU reports GPU properties such as memory size, bandwidth, SM count, and compute capability.

\begin{table}[t]
  \begin{singlespacing}
\centering
\resizebox{\columnwidth}{!}{
  \begin{tabular}{|l|c|r|c|l|}
    \hline
    \textbf{Profiling Rank} & \textbf{KERNEL} & \textbf{MEMCPY} & \textbf{GPU} & \textbf{\# joined} \\
    \hline
    Rank 0 & 842054 & 107045 & 4 & ~93M \\
    \hline
    Rank 1 & 842054 & 107099 & 4 & ~93M \\
    \hline
    Rank 2 & 842054 & 1070545 & 4 & ~93M \\
    \hline
    Rank 3 & 842054 & 107045 & 4 & ~93M \\
    \hline
  \end{tabular}
}
\end{singlespacing}
  \caption{Profiling Ranks and associated SQLITE tables. KERNEL, MEMCPY, GPU is defined as CUPTI\_ACTIVITY\_KIND\_KERNEL, CUPTI\_ACTIVITY\_KIND\_MEMCPY, TARGET\_INFO\_GPU tables respectively. \# joined approximate number of entities after each ``left" joining
  }
  \label{tab:table_for_profiling}
  \vspace{-0.3in}
\end{table}

\section{Design and Implementation}

We design a two-stage pipeline comprising (i) a data generation phase that extracts and shards execution traces, and (ii) a data aggregation phase that consolidates and analyzes the data to surface anomalous behavior. 

\textbf{Data generation}
Our pipeline identifies essential SQLite3 tables and extracts kernel timestamps to define dataset boundaries. We evenly partition the full time range into $N$ non-overlapping shards, each binning kernel executions by timestamp. Given $P$ MPI ranks, we choose block partitioning over cyclic partitioning because the dataset is static and workload predictability is high. Block partitioning assigns contiguous shards to each rank, reducing query overhead, improving data locality, and enabling efficient SQL query execution. Each rank independently processes its assigned shards and saves query results into consistently named parquet files, facilitating seamless downstream aggregation.

\textbf{Data aggregationn}
We begin aggregation by defining a global dictionary with timestamps as keys and a fixed user-defined duration ($interval = 1s$ by default). Each rank loads its assigned $\frac{N}{P}$ parquet files, mapping samples to corresponding time shards. Subsequently, $P$ ranks collaboratively compute statistical metrics (minimum, maximum, standard deviation) in a round-robin manner, balancing workload evenly and minimizing contention. These shared statistics facilitate identification of anomalies, from which we select the top 5 anomalous shards using the Inter-quartile Range (IQR)~\cite{Whaley2014} method.
\section{Preliminary Results}
To evaluate our pipeline, we use the Texas Advanced Computing Center’s (TACC) Lonestar6 supercomputer. The Lonestar6 supercomputer has 560 compute nodes, each with two AMD EPYC 7763 processors (128 cores total) and 256 GB DDR4 memory. There are 84 GPU nodes, each with the same CPUs plus three NVIDIA A100 GPUs (40 GB HBM2 each)



%
\textbf{Memory Stall Duration for all ranks }
Figure~\ref{fig:memory_stall_ranks} plots memory stall durations across all four ranks, with stall durations on the y-axis and elapsed runtime (in seconds) on the x-axis. The plot reveals sustained memory stalls co-occurring across multiple ranks, suggesting synchronization issues or memory bandwidth contention. Based on visual inspection, we select Rank 2 for detailed analysis.

\textbf{Memory Stall vs Kernel Execution Relationship }
We isolate the top 5\% highest-variability intervals from Rank 2 to analyze the relationship between memory stalls and kernel executions (Figure~\ref{fig:memory_stall}). The figure confirms Device-to-Host and Host-to-Device transfers dominate, suggesting frequent ping-pong patterns caused by inefficient batching. In contrast, sparse Device-to-Device transfers indicate infrequent intra-GPU operations, highlighting opportunities for optimization through shared memory reuse or tiling.


\textbf{Overhead}
Figure ~\ref{fig:timing} compares the durations of two phases: Data Generation and Data Aggregation, across different MPI configurations. We find out that both the data generation and aggregation time decrease with respect to increasing the number of MPI ranks. That proves that our pipeline is 
scalable to handle large amount of data.




\section{Conclusion}

We developed a distributed framework that concurrently analyzes GPU performance logs by partitioning SQLite3 tables into shards, reducing memory usage and latency. Analyzing 93M samples, we identified timestamps and entities responsible for memory stalls. Future work will focus on improving scalability by eliminating intermediate I/O bottlenecks.

\section{Acknowledgement}
This material is based upon work supported by the U.S. Department of Energy, Office of Science under Award Number DE-SC0023173. Sandia National Laboratories,  (NTESS) is operated for the U.S. Department of Energy’s National Nuclear Security Administration (DOE/NNSA) under contract DE-NA0003525. Authors thank the Texas Advanced Computing Center (TACC) at The University of Texas at Austin.
\bibliographystyle{ACM-Reference-Format}
\bibliography{sample-base}

\end{document}